\documentclass[conference]{IEEEtran}
\IEEEoverridecommandlockouts
\usepackage{cite}
\usepackage{amsmath,amssymb,amsfonts}
\usepackage[ruled,longend,boxruled]{algorithm2e}
\usepackage{graphicx}
\usepackage{textcomp}
\usepackage{xcolor}
\usepackage{listings}
\def\BibTeX{{\rm B\kern-.05em{\sc i\kern-.025em b}\kern-.08em
    T\kern-.1667em\lower.7ex\hbox{E}\kern-.125emX}}

\begin{document}

\title{A Distributed Partitioning Software and its Applications}

\author{
Aparna Sasidharan\\
Department of Computer Science,\\
\textit{University of Illinois, Urbana-Champaign}\\
IL, USA\\
\textit{aparnasasidharan2017@gmail.com}
}

\maketitle

\begin{abstract}
This article describes a geometric partitioning software that can be used for quick computation of data partitions on many-core HPC machines. It is most suited for dynamic applications with load distributions that vary with time. Partitioning costs were minimized with a lot of care, to tolerate frequent adjustments to the load distribution. The partitioning algorithm uses both geometry as well as statistics collected from the data distribution. The implementation is based on a hybrid programming model that is both distributed and multi-threaded. Partitions are computed by a hierarchical data decomposition, followed by data ordering using space-filling curves and greedy knapsack. This software was primarily used for partitioning 2 and 3 dimensional meshes in scientific computing. It was also used to solve point-location problems and for partitioning general graphs. The experiments described in this paper provide useful performance data for important parallel algorithms on a HPC machine built using a recent many-core processor with large on-chip memory. 
\end{abstract}

Databases, Adaptive Mesh Refinement, Mesh Partitioning, Space-filling Curves, Kd-trees, Intel KNL\\	

\section{Introduction}

Most algorithms in both scientific computing and data processing domains perform read/write operations on data stored in memory, insert new data or remove existing data. But while most numerical methods in scientific computing are based on in-memory matrix algebra, query processing applications may need to retrieve data from disks and access random memory addresses. A parallel partitioner that generates good quality partitions is beneficial to both domains. Our parallel geometric partitioner based on space-filling curves~(SFC) and its performance on Intel KNL many-core processors are discussed in detail in~\cite{sfcthesis}. This paper extends the scope of the partitioner from one that load balances Adaptive Mesh Refinement~(AMR) applications to that of a parallel geometric partitioner that produces good quality partitions for applications ranging from parallel query processing on dynamic workloads to large relationship graphs derived from the internet. Geometric partitioners are rarely used outside structured AMR meshes~\cite{amr}. In this work we describe methods for load balancing unstructured meshes refined using Delaunay methods along with parallel query processing. Geometric partitioning can be applied to general graphs after embedding vertex attributes in $D$-dimensional unit space, where $D\in R$, and defining distance criteria and resolutions for each attribute. Graphs can also be partitioned by partitioning their adjacency matrices as 2D meshes. Our partitioning algorithm assumes that the entire set of co-ordinates fit in the memories of all processes. We provide parallel query processing algorithms such as exact point location and k-nearest neighbors which use space-filling curves. It is important to keep partitioning costs low, because it is an overhead in a parallel algorithm which did not exist in its sequential version. An expensive partitioning algorithm will increase the total work and reduce the speedup of the overall application. Because most HPC applications are hybrid, i.e. distributed and multi-threaded, our partitioning algorithm is also hybrid. Its computation costs are comparable to parallel sorting in the best case. We proved using our AMR implementation that a fast parallel partitioner reduces total execution time by accomodating more load balancing steps and reducing total load imbalance~\cite{sfcthesis}.
The default sorting criterion used by the partitioner is Euclidean distance. The partitioner requires unique global ids for all elements in the input dataset. The output produced is a permutation of these global ids that is partitioned and stored across processing elements. It is left to the application to re-order the dataset according to the partitioner's output.

We improved the quality of geometric partitions by considering the distribution of points in space along with the geometry of the domain, and by defining Hilbert-like SFCs which have better spatial locality. These partitions were compared for load balance and communication volume to those produced by linear optimization~\cite{sfcthesis}. There are several software packages available for meshing and load balancing in the HPC community~\cite{paramesh}, ~\cite{samrai},~\cite{grace},~\cite{changa},~\cite{Chombo},~\cite{enzo},~\cite{FLASH},~\cite{barneshut}. A detailed discussion of related work can be found in ~\cite{sfcthesis}.
The partition problem is discussed in section~\ref{partP} and software architecture in section~\ref{sa}. Section~\ref{ap} discusses partitioning and load balancing of dynamic applications. Applications from different domains that could use our partitioner are described in section~\ref{appl}.

\section{The Partition Problem}\label{partP}

The balanced graph partition problem known to be NP-complete, can be formally defined as: given a graph $G$ with vertex-set $V$ and edge-set $E$, both weighted, a $P$-way partition of the graph should create $P$ disjoint subsets minimizing the maximum weight of a partition or total edge-cut (communication volume) or maximum edge-cut. This is often formulated as an optimization problem with an objective function and a set of constraints. A commonly used objective function is minimization of maximum edge-cut or communication volume, with constraints placed on the maximum load imbalance across partitions. Complex objective functions that minimize the maximum in degree/out degree of partitions or the maximum weighted sum of communication volume and in degree/out degree may be formulated. A couple of different formulations for this problem can be found in Metis~\cite{Metis}. This software also provides options for defining new objectives and constraints.  
The objective function that minimizes maximum communication volume, subject to load imbalance constraints is discussed below. Let $e_i$ be the sum of weights of outgoing edges of any partition $p_i$, which contributes to the total communication volume. The objective function can be formulated as:

\begin{equation}\label{eq:eq1}
\min_{i=1}^{P} \max e_i
\end{equation}

For a given partition set, let $w_i$ be the load~(sum of weights of elements) of any partition $p_i$. Define load imbalance as the maximum difference between the weights of any two partitions $p_i$ and $p_j$. The R.H.S in the constraint is the maximum desired value for load imbalance, say $X$.

\begin{equation}\label{eq:eq2}
\max_{i=1,j=1}^{P} (w_i-w_j) \leq X
\end{equation}

Approaches to solve the partition problem are broadly classified into Spectral Methods~\cite{spectral}, Approximate methods~\cite{approx}, Graph coloring~\cite{color}, Combinatorial Optimization~\cite{kl},~\cite{FM}, Multi-level methods~\cite{kway}, Geometric~\cite{Gilbert_geom}, ~\cite{Samet}, and Streaming Algorithms~\cite{msr_st}.
Most implementations of these methods are sequential. Multi-level methods are widely used in scientific computing for which parallel versions are provided in Parmetis~\cite{parmetis}. There are some recent parallel implementations of spectral methods for small numbers of partitions~\cite{nvidia_spectral}. Graph partitioning software can also be used for partitioning meshes by partitioning their dual graphs. Adaptive meshes may also use partition refinement schemes such as Diffusion~\cite{diffusion} to adjust minor differences in load balance and communication volume. 
Compared to our partitioning algorithm, Parmetis~\cite{parmetis} performs more inter processor communication, is not multi-threaded and performs poorly on many-core processors. 
Recently, there has been a lot of work on developing packages for processing large real-world graphs, typically derived from social networks and web graphs. These are random graphs that follow the power law degree distribution~\cite{snap}. Most of these packages use hash functions to map vertices to bins. A set of bins comprising a partition, were assigned to processes and threads. Since random permutations of vertices were mapped to bins, although load balanced, the partitions had high communication volume and performed poorly. They were later replaced by graph partitioners such as Metis and Parmetis~\cite{Metis}.
Giraph++~\cite{giraph}, GraphX~\cite{graphx}, Dryad~\cite{dryad}, Naiad~\cite{naiad}, DistGraphLab~\cite{distgraphlab}, Mizan~\cite{mizan} and Pregel~\cite{pregel} are some of the widely used packages for graph analytics on real-world data. Naiad uses 2D SFCs to partition adjacency matrices (edges) of graphs. Although SFCs can produce good partitions of adjacency matrices, the performance of an application depends on how those partitions are used. 
There are expensive linear algebra algorithms that re-order and partition adjacency matrices of graphs. Commonly used matrix reordering algorithms are nested-dissection~\cite{nested},~\cite{parallelnest} and reverse cut-hill~\cite{RCM}. \cite{nestedfast} creates coarse partitions of matrices using geometric methods with nested-dissection at the lower levels. 

\section{Software Architecture~\label{sa}}
The parallel partitioning algorithms discussed in this article are suitable for both many-core processors and GPUs. MPI~\cite{mpi} was used for inter process communication and the STL interface to pthreads for multi-threading~\cite{stl}. STL provides APIs for spawning and joining threads, detached thread execution, memory consistency models and synchronization primitives such as locks, barriers and atomic instructions. Thread scheduling is performed by the operating system. The programmer can choose a memory consistency model such as the relaxed memory model that was used by this software\cite{relaxed}. Memory fences were inserted at synchronization points in the program where it was necessary to flush most recent values from local caches to memory. The programs followed SIMD style, with few synchronization points and critical sections. The critical sections were executed by thread 0, while other threads waited for thread 0 to exit the critical section. The implementation is scalable on many-core nodes due to the following reasons :

\begin{enumerate}
\item Parallel Algorithm Design : All the partitioning algorithms used by our implementation had low computation costs. For $n$ points and $p$ processes, the implementation has $O(\frac{n}{p}log\frac{n}{p})$ computation cost if using midpoint splitters, which is optimal for this problem.      
\item Low overhead synchronization : Atomic instructions provided by STL such as fetch-add and compare-swap were used to co-ordinate threads. 
\item Nondeterministic and wait-free algorithms : Some sections of the software were allowed to be non-deterministic without affecting correctness. Allowing non-determinism in the primary data structures reduced program dependencies. The operations on these data structures are linearizable and wait-free, i.e., any thread could progress, but when a thread made progress, it facilitated the progress of blocked threads~\cite{herlihy1}.     	
\end{enumerate}

The software design discussion is divided into three phases :

\begin{itemize}
\item Hierarchical Domain Decomposition
\item SFC Traversal
\item Load Balancing 	
\end{itemize}

\subsection{Hierarchical Domain Decomposition - Kd-trees}

Our tree construction algorithm is recursive, where each recursive step splits a set of points into two subsets and constructs tight bounding boxes around these subsets. Recursion is terminated when the number of points in a subset falls below $BUCKETSIZE$. For meshes, tree construction is independent of the shape of mesh elements. Representative points such as the co-ordinates of the center of gravity were used for partitioning. Elements are indivisible, i.e. all the nodes, faces and edges of a mesh element reside on the same partition. During recursion a node in the tree is divided into exactly two sub cells using a splitting hyperplane in $d-1$ dimensions. Two variables are used to define a splitting hyperplane - splitting dimension and value. For constructing balanced trees, the splitting dimension chosen is always that of maximum width and the value is either the midpoint or median along that dimension. If the splitting dimension is $i$ and value is $m$, then, all points with co-ordinate values less than or equal to $m$ along $i$ are assigned to the $lower$ sub cell and the remaining points to the $upper$ sub cell. Nodes are assigned unique ids and store their splitting hyperplanes. The choice of hyperplanes affect the maximum depth of the kd-tree, its size~(number of nodes) and the time taken for tree construction~\cite{Bentley}.

\begin{figure}[!tbph]
\centering
\includegraphics[width=2.75in,height=2.25in]{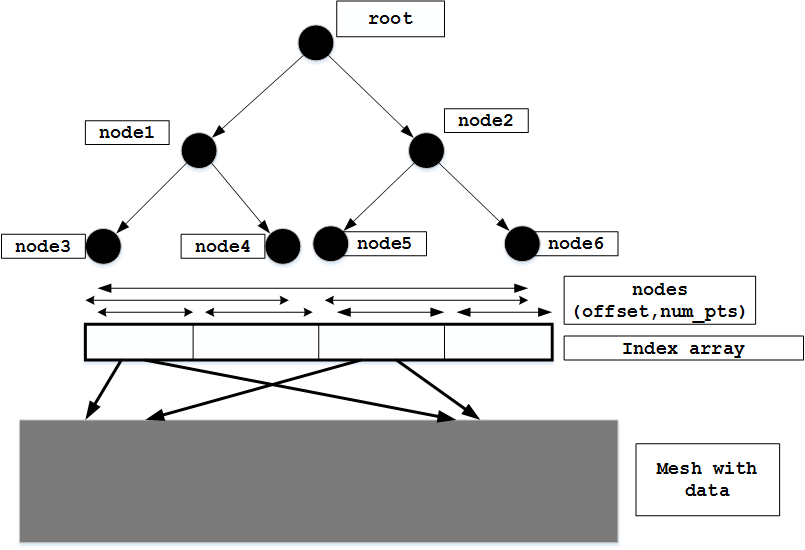}
\caption{\label{Fig:kdtreep}Linearized kd-tree}
\end{figure}

Non-deterministic concurrent linked lists were used to store the tree nodes. Each linked list node is a vector of tree nodes. Atomic variables were used to store link pointers. Threads and processes built different sections of the tree in parallel without any communication and updated a common distributed data structure which is the full kd-tree. Since the addition of tree nodes to the linked list is non-deterministic, different storage orders for nodes are produced in each execution. However, the concurrent data structure is linearizable and sequentially consistent.
Besides tree nodes, the current state of the partitioner was stored in two vectors which are smaller than the original dataset. This improved tree-building time by reducing the total size of memory accessed during partitioning and by improving cache reuse as shown in figure~\ref{Fig:kdtreep}. A vector of indices and a vector of co-ordinates contain the current snapshot of the tree. The input to the program is $N$ points each with $d$ co-ordinates, one unique id, and one weight value, along with $N$ unsigned integers, containing the ids of points.
The implementation supports the following splitting hyperplanes :

\begin{enumerate}
\item Midpoint of the dimension of maximum spread : Geometric midpoint computed by determining the mean of minimum and maximum co-ordinate values along the dimension of maximum width.
\item Exact Median of the dimension of maximum spread  : Median computed by sorting the subset of co-ordinates along a dimension and choosing the middle value.
\item Approximate Median of the dimension of maximum spread  : Median computed by sorting a random subset of co-ordinates in a dimension and choosing the middle value.
\item Approximate Median by Selection : Median computed by ranking a random subset of co-ordinates in a dimension and choosing the value with the median rank.
\end{enumerate}

Details of the distributed and shared memory implementations are discussed here~\cite{sfcthesis}.
Median splitters produced balanced trees for all point distributions at the cost of increased computation. If points are uniformly distributed, midpoint splitters are as good as median splitters for producing balanced trees. For clustered distributions, median splitters produced shorter trees that reduced both tree building and computation times for the operations performed on tree data. A combination of splitters may be used, with median splitters at the top nodes and midpoint splitters at the lower nodes of the tree, to reduce total execution time.
The implementations are divided into two versions, based on the nature of input data - static and dynamic.

\subsubsection{Static Kd-tree~\label{skd}}
For static datasets, the tree once constructed, is maintained in its entirety until the program terminates. Kd-trees built using static datasets can be made space efficient by storing only terminal nodes. Implementations discussed in this article store non-terminal tree nodes for all datasets. 
The static kd-tree is built using $PE$ processing elements, $P$ processes and $T$ threads, where $PE = P*T$ by invoking the following procedures in our library shown in listing~\ref{treebuild}:

\begin{lstlisting}[basicstyle=\scriptsize,caption={Functions for Dynamic Kd-Tree building},captionpos=b,label={treebuild}]
void partitioner_init(m_thread_param *,
        point_d *, unsigned long int);

unsigned long int* point_order_dist_kd(m_thread_param *,
                   void(*splitter1)(m_thread_param *));

unsigned long int* point_order_local_subtree(m_thread_param *,
                   void(*splitter1)(m_thread_param *),
                   void(*splitter2)(m_thread_param *));
\end{lstlisting}

The $partitioner\_init$ function initializes the concurrent data structures necessary for building the tree. The routine $point\_order\_dist\_kd$ initializes and traverses the top $K1$ nodes of the tree, where $K1>=P$. This section of the implementation is distributed across multiple processes and requires inter process communication for computing splitters. Every top node in the tree has a unique SFC key assigned to it during tree traversal. The generation of SFC keys for Morton and Hilbert-like curves are explained in detail in~\cite{sfcthesis}. After the entire dataset is assigned membership to one of the top $K1$ nodes, these nodes are re-ordered according to their SFC keys and assigned to processes. Node weights are equal to the sum of weights of points in them and a greedy knapsack function assigns roughly equal weights to processes. The points of the data set are reordered according to partitions of top nodes. For a pair of processes $P_i$ and $P_j$, where $i < j$, all nodes assigned to $P_i$ have SFC keys strictly less than those assigned to $P_j$. The second routine $point\_order\_local\_subtree$ is executed locally by processes. Sub trees are built in two phases. The first phase builds the top $K2$ nodes of local sub trees. The $K2$ nodes, $K2>=T$ are built in breadth-first order, assigned SFC keys and partitioned across $T$ threads using greedy knapsack. In the second phase, threads work independently by constructing their sub trees in depth-first order.

\begin{figure}[!tbph]
\centering
\includegraphics[width=4in,height=3in]{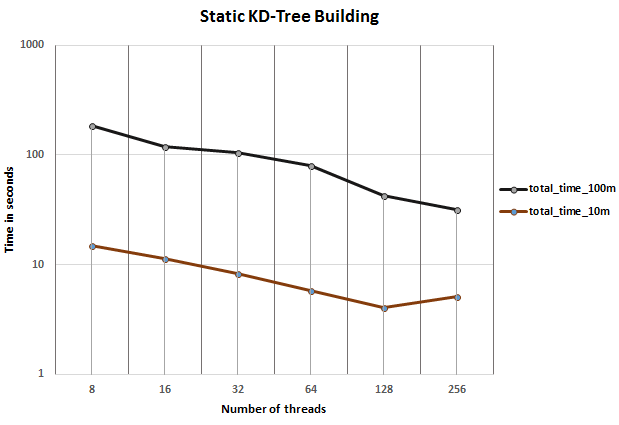}
\caption{\label{Fig:static_strong}Static KD-tree, strong scaling, uniform distribution, midpoint splitter}
\end{figure}

Performance results are presented for shared memory and distributed memory implementations separately.
Strong scaling results for kd-tree construction on a single node are presented. The test cases used were uniform point distributions with 10 million and 100 million points in 3D. Midpoint splitters were used for constructing the trees. The number of threads per process were varied from $8$ to $256$. All tests were performed using Intel KNL nodes configured to use the high bandwidth memory~(MCDRAM) as L3 cache. There are 34 functional tiles, 68 cores and 272 hardware threads per KNL node. Out of these resources, at most 32 tiles and 256 threads were used. Results are shown in the graph in ~\ref{Fig:static_strong}. The y-axis of this graph is logarithmic scale.

Two kinds of test cases were used for evaluating the static kd-tree and choice of splitting hyperplanes. The first test case used a uniform point distribution~\cite{stlrandom}. Experiments were performed using different thread counts and problem sizes. The measured values are averaged over five runs. Buckets sizes were fixed at $32$ for all test cases except $100$ million points. The bucket size for $100$ million points was $128$.

\begin{figure}[!tbph]
\centering
\includegraphics[width=3.2in,height=2.2in]{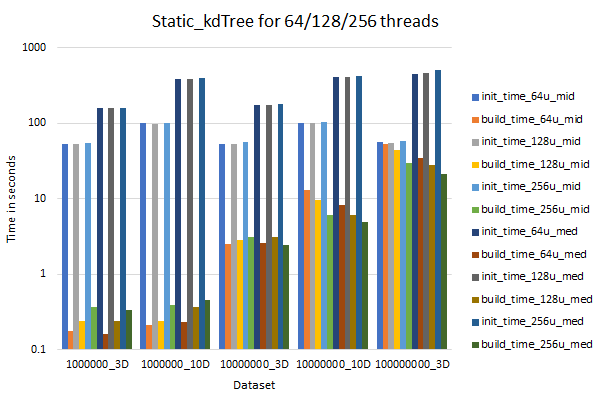}
\caption{\label{Fig:static_uniform}Static Kdtree, Uniform Distribution, Median Splitter~(Sorting)}
\end{figure}

\begin{figure}[!tbph]
\centering
\includegraphics[width=3.2in,height=2.2in]{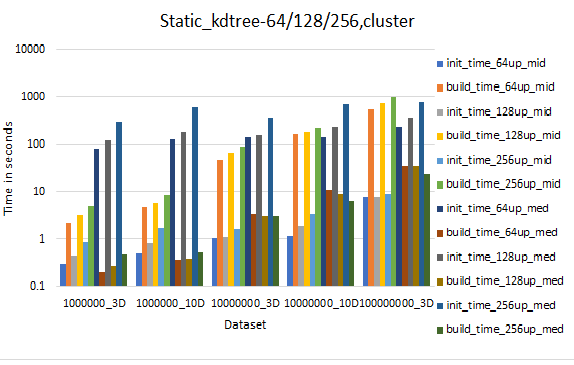}
\caption{\label{Fig:static_cluster1}Static Kdtree, Cluster, Median Splitter~(Sorting) }
\end{figure}

\begin{figure}[!tbph]
\centering
\includegraphics[width=3.2in,height=2.2in]{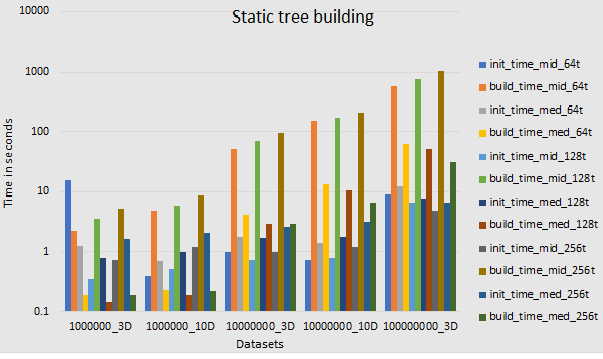}
\caption{\label{Fig:static_cluster2}Static Kdtree, Cluster, Median Splitter~(Selection)}
\end{figure}

The second test case was a clustered distribution, created by mixing a Poisson distribution with mean value in the bottom left corner of a hypercube domain and a uniform distribution. For the clustered distribution, tree building times using midpoint splitting hyperplanes were high because of unbalanced trees. The differences in tree building times between midpoint and median splitting hyperplanes are apparent in the results in this section.
For the graphs in ~\ref{Fig:static_uniform} and ~\ref{Fig:static_cluster1}, the $partitioner\_init$ phase with median splitters was expensive. Median values were computed by sorting co-ordinates along the splitting dimensions and by choosing middle values. Sorting was performed using a distributed concurrent quick sort implementation~\cite{sfcthesis}. For large datasets, $point\_order\_local\_subtrees$ times with median splitters were lower than trees with midpoint splitters. The improved execution times with median splitters with selection instead of sorting are shown in figure ~\ref{Fig:static_cluster2}.

\subsection{Space-filling Curve~(SFC) Traversals~\label{sfc}}

\begin{figure}[!tbph]
\centering
\includegraphics[width=3in,height=3in]{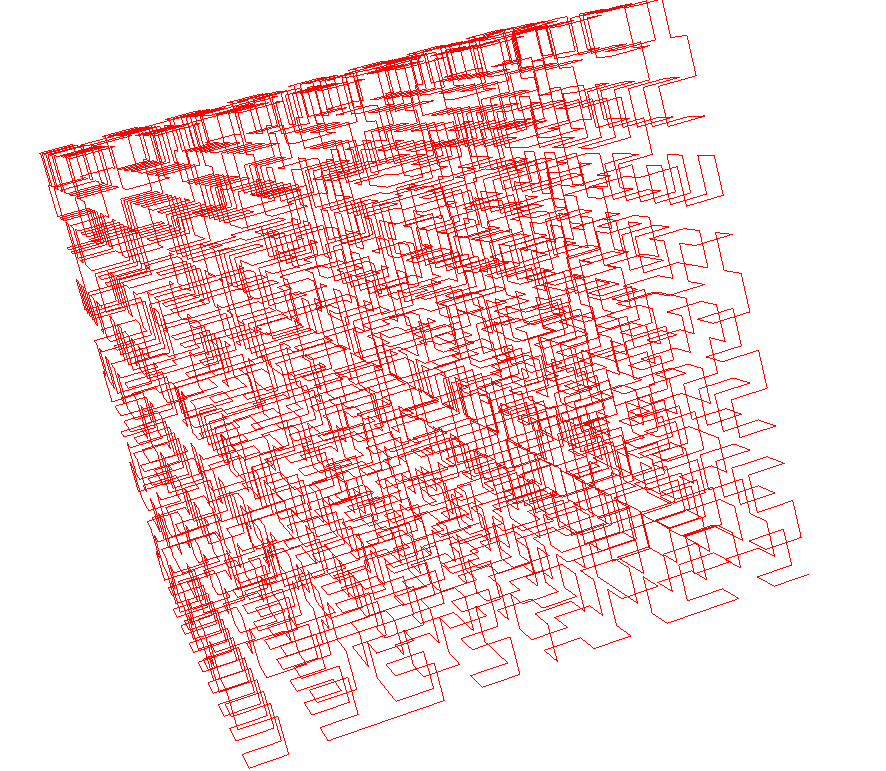}
\caption{\label{fig:sfc_3d_reg} A 3D Hilbert-like Curve}
\end{figure}

\begin{figure}[!tbph]
\centering
\includegraphics[width=3in,height=3in]{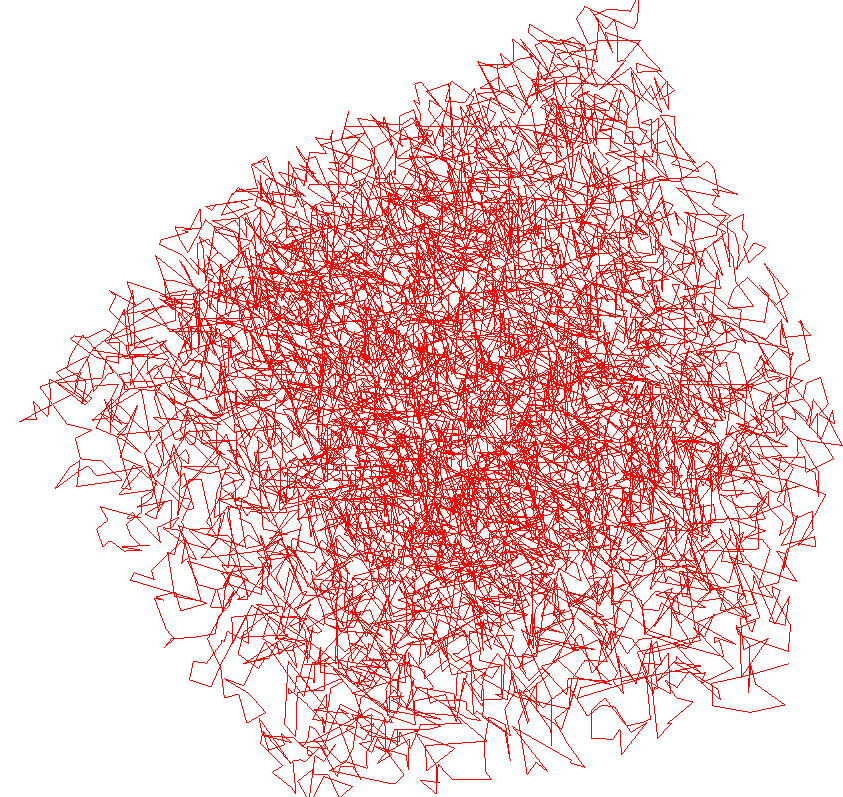}
\caption{\label{fig:sfc_3d_irr} A 3D Hilbert-like Curve on Irregular Distribution}
\end{figure}

Once trees are built, they are traversed from top nodes to leaves to assign SFC keys. At the end of this the global ids of points are stored in the sorted order of SFC keys. Two space-filling curves are supported by this partitioner. The default SFC is Morton~\cite{Morton}. A Hilbert-like~\cite{Sagan} curve is provided by the partitioner that extended the geometric definition of Hilbert curves to include random point distributions and unstructured meshes, shown in figures~\ref{fig:sfc_3d_reg} and \ref{fig:sfc_3d_irr}. Both Morton and Hilbert-like curves are recursive constructions that order points based on the order of traversal of tree nodes. SFC traversals are relatively cheap operations compared to tree building. Increase in the number of dimensions increases the degrees of freedom in the curve and its geometric transformations. Our implementation has no restrictions on the number of dimensions. Hilbert-like curves are generated recursively during traversals using a set of rules for the visiting order of sub cells. Base rules are defined for 2D and extended to higher dimensions by repetition and concatenation~\cite{sfcthesis}.

\begin{figure}[!tbph]
\centering
\includegraphics[width=8cm, height=6cm]{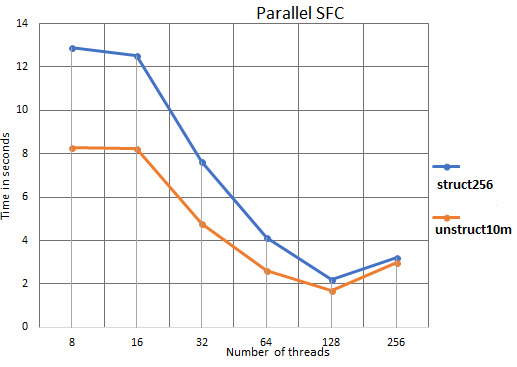}
\caption{\label{fig:parallelsfc1} Parallel SFC on 256X256X256 mesh and 10m points, single-node performance}
\end{figure}

\begin{figure}[!tbph]
\centering
\includegraphics[width=7cm,height=5cm]{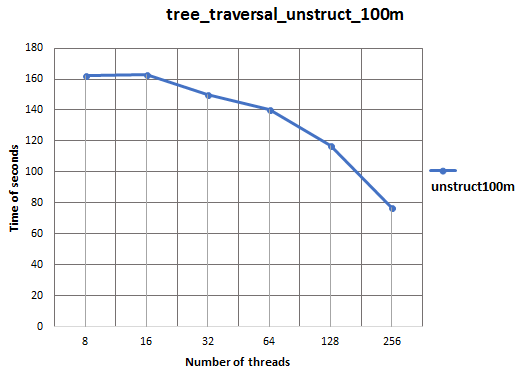}
\caption{\label{fig:parallelsfc2} Parallel SFC on 100m points, single-node performance}
\end{figure}

Our Morton and Hilbert-like traversals are parallel implementations. Unlike Morton, the Hilbert-like traversals require look-ahead during tree traversals, which result in minor increase in traversal times. But the SFCs produced by Hilbert-like curves have better spatial locality which results in partitions with lower surface to volume ratios. For a given number of points in a partition, its communication volume is equal to the weighted sum of its outgoing edges. Good quality partitions with load balance and low maximum communication volume are beneficial for iterative algorithms which involve several iterations of computation and inter process communication with nearest neighbors. All measurements reported in this section are the total times which includes both tree building and Hilbert-like SFC traversals.
Strong scaling results for Hilbert-like traversal are shown in figures ~\ref{fig:parallelsfc1} and ~\ref{fig:parallelsfc2}. All experiments in this section were performed on Stampede2~\cite{Stampede2}. Details of this machine are provided in section~\ref{ap}. 
Figure~\ref{fig:parallelsfc1} shows the performance for a regular mesh of dimensions $256X256X256$ and a random distribution of 10 million points. Both test cases used $BUCKETSIZE=32$. Figure~\ref{fig:parallelsfc2} shows the time taken for traversing a random distribution of 100 million points in parallel using a single KNL node.

\begin{figure}[!tbph]
\centering
\includegraphics[width=9cm, height=8cm]{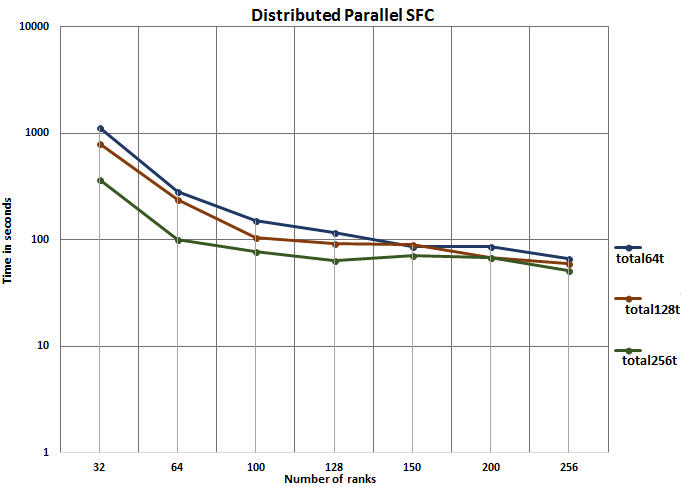}
\caption{\label{fig:parsfc8b} Parallel SFC on a uniform distribution of 8 billion points}
\end{figure}

Figure~\ref{fig:parsfc8b} shows the strong scaling performance of parallel Hilbert-like SFC on 8 billion points. This is a distributed memory implementation also tested on Stampede2.

\subsection{Load Balancing}

SFC traversals store points in the sorted order of SFC keys across processes and threads. For any two processes $P_i$ and $P_j$ where $i < j$, SFC keys on $P_i$ are strictly less than those on $P_j$, and for any two threads $T_{i_i}$ and $T_{i_j}$ on $P_i$ where $i_i < i_j$ keys on $T_{i_i}$ are strictly less than those on $T_{i_j}$. After reordering, points are partitioned using a parallel implementation of the greedy knapsack algorithm~\cite{sfcthesis}. Processes compute their local weights followed by parallel reduction to compute the total weight that is distributed across processes. A parallel prefix computation is used to determine the global rank of a point on a weighted line segment~(SFC) of points. For $P$ processes, this weighted line segment is sliced into $P$ partitions of almost equal weights without violating the sorted order of SFC keys. The load on any two processes differs by at most the maximum weight of any point. 

\begin{lstlisting}[caption={Functions for Data Migration in distributed trees},captionpos=b,label={listdyndist}]

 load_balance(tp);
 transfer_t_l_t(tp);
\end{lstlisting}
The functions for data distribution in distributed trees are provided in listing~\ref{listdyndist}.
The $load\_balance$ routine in listing~\ref{listdyndist} computes the partitions of points and includes tree-building, SFC traversal and greedy knapsack. $transfer\_t\_l\_t$ in listing~\ref{listdyndist} is the data exchange or data migration routine that migrates stored data between processes according to these partitions. Our implementation performs data exchange in rounds, by placing an upper limit on the maximum message size~($MAX\_MSG\_SIZE$). The $transfer\_t\_l\_t$ function packs data into communication buffers, exchanges them using MPI function calls and unpacks received data. Both packing and unpacking routines are concurrent multi-threaded implementations.

\section{Partitioning and Load balancing Dynamic Data ~\label{ap}}

\begin{algorithm}
\small
\DontPrintSemicolon
\caption{Adjustments: Algorithm for sub tree adjustments\label{alg:adjustments}}
\KwIn{node}
\KwOut{double}
\SetKwBlock{Begin}{procedure}{end procedure}
\SetKwFunction{isLeaf}{isLeaf}
\SetKwFunction{LeftChild}{LeftChild}
\SetKwFunction{RightChild}{RightChild}
\SetKwFunction{Adjustments}{Adjustments}
\SetKwFunction{SplitLeaf}{SplitLeaf}
\SetKwFunction{SetLeaf}{SetLeaf}
\SetKwFunction{Parent}{Parent}
\SetKwFunction{SetChild}{SetChild}
\SetKwFunction{newBucket}{newBucket}
\SetKwFunction{SetBucket}{SetBucket}
\Begin(\text{Adjustments}{(}n{)})
{
\If{\isLeaf{n}}
{
\If{$n.wt > 2*BUCKETSIZE$}
{
\SplitLeaf{n}\;
\SetLeaf{n,false}\;
}
\KwRet{n.wt}\;
}
\Else
{
$w1 \gets 0$; $w2 \gets 0$\;
\If{\LeftChild{n}}
{
$w1 \gets \Adjustments{\LeftChild{n}}$\;
\If{$w1=0$}
{
  \SetChild{n,left,NULL}\;
}
}
\If{\RightChild{n}}
{
$w2 \gets \Adjustments{\RightChild{n}}$\;
\If{$w2=0$}
{
\SetChild{n,right,NULL}\;
}
}
$n.wt \gets w1+w2$\;
\If{$n.wt \leq BUCKETSIZE$}
{
$l \gets \LeftChild{n}$; $r \gets \RightChild{n}$\;
\If{$l \land r$}
{
\If{$\isLeaf{l} \land \isLeaf{r}$}
{
  $b \gets \newBucket{}$\;
  $b \gets l.b + r.b$\;
  \SetChild{n,left,NULL}\;
  \SetChild{n,right,NULL}\;		
  \SetLeaf{n,true}\;	
  \SetBucket{n,b}\;
}
}
\Else
{
\If{$l \land \isLeaf{l} \land \neg r$}
{
  $b \gets \newBucket{}$\;
  $b \gets l.b$\;
  \SetChild{n,left,NULL}\;
  \SetLeaf{n,true}\;
  \SetBucket{n,b}\;
}
\ElseIf{$r \land \isLeaf{r} \land \neg l$}
{
   $b \gets \newBucket{}$\;
   $b \gets r.b$\;
   \SetChild{n,right,NULL}\;
   \SetLeaf{n,true}\;
   \SetBucket{n,b}\;
}
}
}

\KwRet{n.wt}\;
}

}
\end{algorithm}

\begin{algorithm}
\small
\DontPrintSemicolon
\caption{Load\_Balancing: Algorithm for Full load balancing\label{alg:lbfull}}
\KwIn{}
\KwOut{}
\SetKwBlock{Begin}{procedure}{end procedure}
\SetKwFunction{BuildTree}{BuildTree}
\SetKwFunction{SFCTraverse}{SFCTraverse}
\SetKwFunction{GreedyKnapsack}{GreedyKnapsack}
\SetKwFunction{ConcurrentAdjustments}{ConcurrentAdjustments}
\Begin(\text{LoadBalance}{(}{)})
{
\BuildTree{}\;
\SFCTraverse{}\;
\GreedyKnapsack{}\;
\ConcurrentAdjustments{}\;
}
\end{algorithm}

\begin{algorithm}
\small
\DontPrintSemicolon
\caption{Dynamic\_Pointset: Algorithm for Amortized Loadbalancing\label{alg:dyptset}}
\KwIn{max\_iter,step\_size,n}
\KwOut{bool}
\SetKwBlock{Begin}{procedure}{end procedure}
\SetKwFunction{BuildTree}{BuildTree}
\SetKwFunction{NewPoints}{NewPoints}
\SetKwFunction{RemPoints}{RemPoints}
\SetKwFunction{Buckets}{Buckets}
\SetKwFunction{UpdateIndex}{UpdateIndex}
\SetKwFunction{LeafTraverse}{LeafTraverse}
\SetKwFunction{DeleteNodes}{DeleteNodes}
\SetKwFunction{SplitNodes}{SplitNodes}
\SetKwFunction{MergeNodes}{MergeNodes}
\SetKwFunction{Adjustments}{Adjustments}
\SetKwFunction{SFCTraverse}{SFCTraverse}
\SetKwFunction{GreedyBinning}{GreedyBinning}
\SetKwFunction{wallTime}{wallTime}
\SetKwFunction{ReduceBcast}{ReduceBcast}
\SetKwFunction{ReduceTh}{ReduceTh}
\SetKwFunction{Insert}{Insert}
\SetKwFunction{Delete}{Delete}
\SetKwFunction{InsertDelete}{InsertDelete}
\SetKwFunction{ThreadId}{ThreadId}
\SetKwFunction{NumThreads}{NumThreads}
\SetKwFunction{LoadThread}{LoadThread}
\SetKwFunction{LoadDistThread}{LoadDistThread}
\SetKwFunction{Spawn}{Spawn}
\SetKwFunction{Join}{Join}
\SetKwFunction{NumBuckets}{NumBuckets}
\SetKwFunction{LoadBalance}{LoadBalance}
\Begin(\text{Dynamic}{(}max\_iter,step\_size,n{)})
{
$t1 \gets \wallTime{}$\;
\LoadBalance{}\;
$t2 \gets \wallTime{}$\;
$totalb \gets \ReduceBcast{\NumBuckets{},MAX}$\;
$lbtime \gets \ReduceBcast{t2-t1,MAX}$\;
$\delta \gets 0$; $basetimeop \gets 0$; $basebkt \gets 0$\;
$n \gets \NumThreads{}$\;
$td \gets \ThreadId{}$\;
\For{$iter \in 1,max\_iter$}
{
\If{$iter\%step\_size=0$}
{
\tcc{Get points for insertion and deletion}
$adlist \gets \NewPoints{n}$ \;
$adlist \gets \RemPoints{n}$ \;
$alist \gets \LoadDistThread{adlist,td}$\;
$ctime \gets 0$\;
\tcc{Insert/Delete points}
$t1 \gets \wallTime{}$\;
$\Spawn{n}$\;
\For{$i \in alist$}
{
$\InsertDelete{alist(i)}$\;          
}
$\Join{n}$\;
$t2 \gets \wallTime{}$\;
$ctime \gets \ReduceBcast{t2-t1,MAX}$\;
$numops \gets \ReduceBcast{alist.size(),SUM}$\;
$timeperop \gets \frac{ctime}{numops}$\;
\If{$basetimeop=0$}
{
  $basetimeop \gets timeperop$\;
  $basebkt \gets basetimeop*totalb$\;
  $\delta \gets 0$\;
}
\Else
{
$timebkt \gets timeperop*totalb$\;
\If{$timebkt > basebkt$}
{
  $\delta \gets \delta+timebkt-basebkt$\;
}
}
\If{$\delta \geq lbtime$}
{
$t1 \gets \wallTime{}$\;
\LoadBalance{}\;
$t2 \gets \wallTime{}$\;
$totalb \gets \ReduceBcast{\NumBuckets{},MAX}$\;
$lbtime \gets \ReduceBcast{t2-t1,MAX}$\;
$\delta \gets 0$;$basetimeop \gets 0$\;
}
}
\If{$iter\%2*step\_size=0$}
{
$nn \gets \LoadThread{topnodes,n}$\;
\Spawn{n}\;
\For{$i \in nn$}
{
$\Adjustments{i}$ \;
}
\Join{n}\;
$totallb \gets \ReduceBcast{\NumBuckets{},MAX}$\;
}
}
}
\end{algorithm}

We use the term \emph{dynamic} in this paper to refer to applications which have variable loads during execution such as AMR. Dynamic applications require multiple load balancing operations to ensure balanced load distributions throughout the execution of the program. But frequent load balancing increases total work which will reduce speedup and adversely affect the scalability of the application. One of the reasons for high load balancing costs is data migration which rearranges the full dataset. The inter processor communication cost of data migration depends on the total communication volume, network topology, hardware and congestion in the network. We introduced amortized load balancing to AMR in our previous work~\cite{sfcthesis}. By minimizing the number of load balancing operations, this technique improved the load balance and reduced the total execution time of AMR simulations. But those simulations were limited to structured AMR with quad tree and oct tree meshes. In this section, we extend our SFC partitioning algorithm and load balancing techniques to dynamic applications such as Delaunay mesh refinement and parallel query processing algorithms on d-dimensional point data.
We used distributed dynamic weighted trees for partitioning and load balancing dynamic data. Leaf nodes are buckets with at most $BUCKETSIZE$ points. We defined \emph{heavy} and \emph{light} buckets, where \emph{heavy} buckets have sizes that exceed $2*BUCKETSIZE$ and \emph{light} buckets have close to zero points. \emph{Heavy} buckets are split recursively into smaller buckets and \emph{light} buckets are merged. These two operations referred to as \emph{adjustments} are described in algorithm~\ref{alg:adjustments}.
The initial weighted kd-tree is built from archived data. If input distributions are clustered, median splitters may be used for building the top $K1*K2*P$ nodes of the tree, where $K1$ and $K2$ are constants and $P$ is the number of processes. Once the top nodes are built and assigned to threads a concurrent implementation of algorithm~\ref{alg:adjustments} is used to compute node weights in sub trees. The algorithm described in algorithm~\ref{alg:adjustments} describes the computation of node weights for a single sub tree. During traversal, this algorithm splits \emph{heavy} buckets and merges \emph{light} buckets. \emph{SplitLeaf} in algorithm~\ref{alg:adjustments} splits leaf buckets recursively until all buckets are within $BUCKETSIZE$. These operations are required for maintaining constant computation cost per bucket and for removing lengthy sub trees with total weight less than $BUCKETSIZE$. SFC keys are updated during splitting and merging operations. Algorithm~\ref{alg:dyptset} describes a full dynamic application with amortized load balancing. We used a dynamic application with explicit queries that executes for a fixed duration to illustrate the load balancing algorithm. This application receives insert/update/delete queries which are distributed to processes based on the partitions of the top $K1*K2*P$ nodes of the tree by \emph{LoadDistThread}. Queries are processed periodically using a fully distributed algorithm. \emph{InsertDelete} processes queries by locating~(depth-first search) and updating buckets. \emph{ReduceBcast} in algorithm~\ref{alg:dyptset} performs global reduction on a vector using a binary operator and \emph{LoadThread} partitions a vector locally between threads. The algorithm described here is iterative, performs computation or processes queries in steps, where the value of $step\_size$ can be adjusted to match the needs of the application. In our extension to the amortized load balancing scheme~\cite{sfcthesis}, the credits accumulated by a load balancing phase are amortized over all the load imbalances in the following iterations. We consider a load balanced computation as incurring zero cost. The next load balancing phase is invoked when all credits are exhausted. In our earlier implementations, we used measured computation time as costs for amortization. Computation cost would work for all iterative applications in scientific computing, such as AMR and Delaunay mesh refinement. We had to modify our definitions of computation cost and load imbalance for query processing applications. We defined computation cost as the product of the maximum average cost per query and the maximum number of buckets across all processes. This quantity can detect load imbalance because it is a measure of the maximum load on any process. In the algorithm, we measured the computation cost after a load balancing phase and monitored its variations in following iterations. Increase in computation cost is paid using credits accrued by the most recent load balancing phase. The next load balancing phase is invoked when all credits are expended. In Algorithm~\ref{alg:dyptset}, $max\_iter$ is the maximum number of iterations until termination.

Load balancing increases the amortized cost for computations in applications. This increase in computation cost depends on the frequency and the cost of each load balancing phase. The amortized cost per operation can be further reduced if incremental load balancing is used instead of full load balancing~(described here). 
Since SFCs preserve spatial locality of data, this method is suitable for incremental load balancing. Our incremental load balancing algorithm which was used for AMR, skips tree building and SFC traversals and recomputes ranks for all points on a new weighted space-filling curve. The greedy knapsack algorithm is used to slice the curve into $P$ almost equal weights. For small changes in load, besides performing less work, our incremental load balancing algorithm has lower inter processor communication because for any process $P_i$, data migration is restricted between $P_i$ and its two neighbors $P_i-1$ and $P_i+1$ in the best case. However, after several iterations of data modifications, the point distribution in the domain is likely to become skewed and the initial statistics used for creating partitions will no longer hold. In such cases, partitions although load balanced are likely to have high surface to volume ratios, which could affect the scalability of mesh refinement applications by increasing inter process communication during nearest neighbor updates. Misshapen partitions can be detected by computing the surface to volume ratios of partitions and the user may switch to a full load balancing to improve partition quality. In algorithm~\ref{alg:dyptset} described here we have used full load balancing. Incremental load balancing will require a distributed version of algorithm~\ref{alg:adjustments}, which performs adjustments. Here we assume that entire sub trees reside on the same process. In order to support incremental load balancing, algorithm~\ref{alg:adjustments} should be modified to include remote parent-child updates.

The discussions and experiments in this section have only considered scenarios where the entire dataset fit in the memories of processes. If datasets are too large to fit in memory, the weighted kd-trees should be external. Pages~($4MB$) should be used instead of in-memory buckets. Demand-paging may be used to read pages from disks and memory and pages have to be managed to reduce the total number of disk accesses.

\subsection{Testcases}

\subsubsection{AMR Performance with Amortized Load Balancing}

\begin{figure}[!tbph]
\includegraphics[width=3.5in,height=3in]{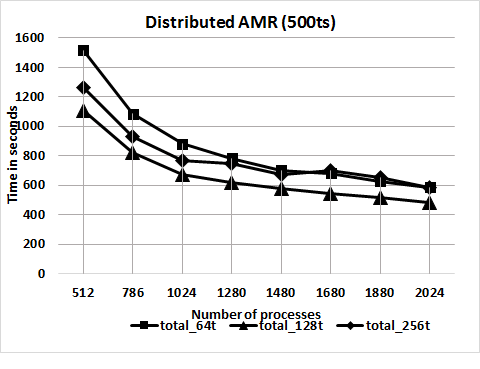}
\caption{\label{fig:amr2}AMR Distributed Memory Performance - Experiment2}
\end{figure}

\begin{table}[!tbph]
\begin{tabular}{|c|c|c|c|c|c|}
\hline
nodes &threads & Refinement & Stencil & LBTime Total\\
\hline
512 & 64 & 816.872 & 271.338 & 114.649 & 1515.91 \\
\hline
786 & 64 & 410.447 & 194.344 & 71.2939 & 1083.35 \\
\hline
        1024 & 64 & 336.064 & 160.537 & 61.7675 & 881.984\\
\hline
        1280 & 64 & 282.92 & 130.879 & 75.1891 & 783.838\\
\hline
        1480 & 64 & 250.918 &  120.501 & 69.7705 & 700.528\\
\hline
        1680 & 64 & 226.035 & 106.726 & 92.6569 & 679.566\\
\hline
        1880 & 64 & 208.684 & 104.613 & 82.9493 & 625.029\\
\hline
        2024 & 64 & 194.558 &  92.9724 & 75.6809 & 584.729\\
\hline
\end{tabular}
\vspace{1mm}
\caption{\label{tab:amr_l64} AMR Thread-local lists,64 threads}
\end{table}

\begin{table}[!tbph]
\begin{tabular}{|c|c|c|c|c|c|}
\hline
nodes &threads & Refinement & Stencil & LBTime Total\\
\hline
        512 & 128 & 596.071 & 223.691 & 85.1687 & 1103.35 \\
\hline
        786 & 128 & 298.707 & 163.363 & 75.2407 & 818.507 \\
\hline
        1024 & 128 & 245.666 & 132.52 & 64.2092 & 674.458\\
\hline
        1280 & 128 & 209.238 & 115.055 & 82.1405 & 619.831\\
\hline
        1480 & 128 & 196.709 &  101.826 & 80.869 & 577.016\\
\hline
        1680 & 128 & 165.648 & 90.677 & 94.0432 & 542.8\\
\hline
        1880 & 128 & 157.005 & 87.8898 & 96.9587 & 516.375\\
\hline
        2024 & 128 & 138.923 &  77.8763 & 87.6129 & 485.262\\
\hline
\end{tabular}
\vspace{1mm}
\caption{\label{tab:amr_l128}AMR Thread-local lists, 128 threads}
\end{table}

\begin{table}[!tbph]
\begin{tabular}{|c|c|c|c|c|c|}
\hline
nodes &threads & Refinement & Stencil & LBTime Total\\
\hline
        512 & 256 & 701.498 & 246.468 & 113.614 & 1260.58 \\
\hline
        786 & 256 & 339.834 & 178.717 & 108.747 & 927.069 \\
\hline
        1024 & 256 & 278.381 & 143.177 & 100.76 & 764.855\\
\hline
        1280 & 256 & 250.572 & 126.017 & 132.481 & 746.569\\
\hline
        1480 & 256 & 220.025 &  113.748 & 138.08 & 675.061\\
\hline
        1680 & 256 & 230.386 & 106.852 & 128.422 & 701.121\\
\hline
        1880 & 256 & 196.446 & 102.838 & 131.345 & 651.716\\
\hline
        2024 & 256 & 173.745 &  96.1577 & 114.074 & 586.495\\
\hline
\end{tabular}
\vspace{1mm}
\caption{\label{tab:amr_l256} AMR Thread-local lists, 256 threads}
\end{table}

\begin{figure}[!tbph]
\includegraphics[width=3.5in,height=3in]{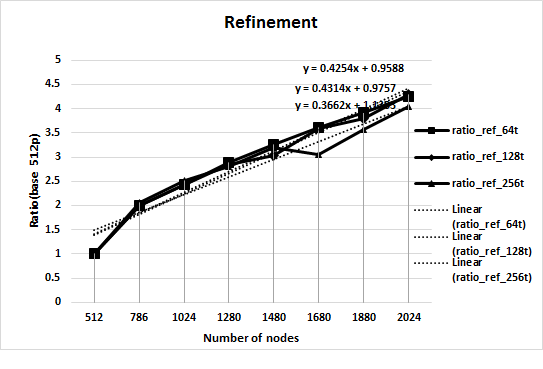}
\caption{\label{fig:ref2}Refinement Distributed Memory Performance -Experiment2}
\end{figure}

\begin{figure}[!tbph]
\includegraphics[width=3.5in,height=3in]{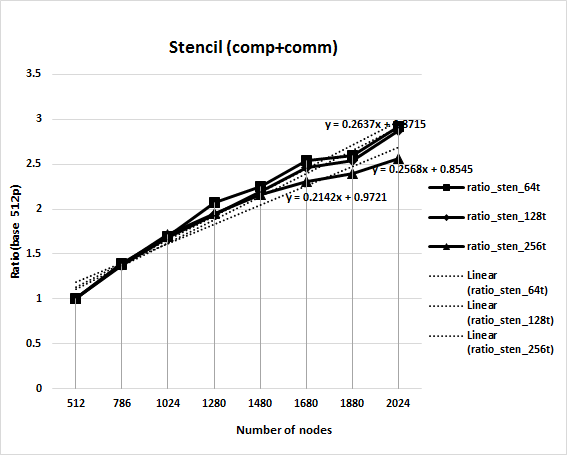}
\caption{\label{fig:sten2}Stencil Distributed Memory Performance - Experiment2}
\end{figure}

This section describes the performance of our AMR benchmark with amortized load balancing, since it was the motivating problem for this partitioner.
Since our AMR algorithms and implementation are described in detail in \cite{sfcthesis}, here we provide observations from a large mesh refinement experiment with many stages. It demonstrates the capability of the partitioner in handling large volumes of dynamic data distributed across thousands of many-core nodes. This is a rapidly evolving test case which requires scalable distributed algorithms to avoid overheads from load imbalances. The initial mesh had $200X200X200$ blocks~(20 billion points). The setup was a small spherical object that moved from the lower left corner to the upper right corner of the mesh domain. There were refinements at almost every iteration. Because of highly parallel AMR algorithms which led to better utilization of cores, low computation overheads and amortized load balancing we could obtain strong scaling performance for this experiment. The observations are tabulated in the tables~\ref{tab:amr_l64},~\ref{tab:amr_l128} and ~\ref{tab:amr_l256}. The graph in figure~\ref{fig:amr2} shows the total time taken for $500$ timesteps versus the number of KNL nodes. Strong scaling can be observed in the speedup graphs in figures~\ref{fig:ref2}~(refinement) and ~\ref{fig:sten2}~(stencil). The baseline for computing speedup was the performance on 512 processes. We have separated the speed ups for mesh refinement and stencil computation because they are different algorithms and it will be useful for analysing hotspots in their implementations. The number of KNL nodes were varied from $[512-2024]$. One MPI process was placed per node. $64$ cores were used per node and $[1-4]$ threads were used per core. The total number of threads used in this experiment ranged from $512*64$ to $2024*256$.

\subsubsection{General Dynamic Applications}

Evaluation of algorithm~\ref{alg:dyptset} is described in this section. 
New points were created by sampling from the domain bounding box. For test cases in this section, new points were sampled every 100 iterations and inserted into the tree. Adjustments were performed every 500 iterations. The dynamic tree-building program was executed for a maximum of 1000 iterations. All experiments were carried out on the Stampede2 supercomputer~\cite{Stampede2} at the Texas Advanced Computing Center~(TACC). As of 2024, Stampede2 had 4,200 Intel Knights Landing (KNL) nodes and 1,736 Intel Xeon Skylake nodes. But the results published in this paper are from 2018, when the machine had about 1000 Intel KNL nodes. The communication network on the machine uses a 100 Gb/s Intel Fabrics Division Omni-Path network and has fat-tree topology.  A single Intel KNL node was used for measuring the performance of shared memory implementations.
Results are tabulated for initial datasets of sizes 1 million and 10 million points in 3D and 10D. $BUCKETSIZE$ was $32$ for all test cases with 1 million points. For 10 million points, $BUCKETSIZE$ was 100. The reported measurements are accumulated over 1000 iterations. The results provided in table~\ref{tab:dykdtree} are for a uniform distribution with midpoint splitters.

\begin{table*}[!tbph]
\small
\begin{center}
\begin{tabular}{|c|c|c|c|c|c|c|c|}
\hline
        \#th & points & nodes & build & ins &del&adj&total\\
\hline
        64 & 1m3D & 90771 & 1.0326 & 0.25673 & 0.901217&3.9307441&78.8781 \\
\hline
        128 & 1m3D & 90909 & 1.60241 & 1.39667 & 0.185382&0.714679& 16.4273522\\
\hline
        256 & 1m3D & 90853 & 2.79706& 3.32358 & 0.223829   &1.74972& 36.03697506 \\
\hline
        64 & 1m10D & 94823 & 3.35145& 1.13261 & 0.230857&   0.850745&5.7140883\\
\hline
        128 & 1m10D & 94577 & 3.97817&1.32123 & 0.162733 &0.71896&17.84933349\\
\hline
        256 & 1m10D & 94731 &6.06864&2.76369 &0.238967 &1.15153&47.4207711\\
\hline
        64 & 10m3D & 289371 & 24.6541&15.1704 &3.61651 &16.9726&61.04216976 \\
\hline
        128 & 10m3D & 289339 & 20.3154&17.6134 &2.22621 &15.591&87.5369396\\
\hline
        256 & 10m3D & 289737&23.7506&40.6131&2.2948&26.7047&164.1104614\\
\hline
        64 & 10m10D & 314629& 52.9961&15.7457&4.19934&18.4795&91.9965334\\
\hline
        128 & 10m10D & 315361&58.2669&20.1613&2.85784&14.8437&129.6039031\\
\hline
        256 & 10m10D & 315277& 73.2034&47.3685&2.81898&26.0353&226.457175\\

\hline
\end{tabular}
\vspace{1mm}
\caption{\label{tab:dykdtree}Dynamic KD-tree construction time, midpoint splitter}
\end{center}
\end{table*}

These results do not show strong scaling for all data sizes. One of the problems with managing dynamic data is cache misses. Changing data sizes, adjustments and load balancing lead to re-assignment of sub trees to threads and causes cache misses. This resulted in reduced performance for $128$ and $256$ threads for some datasets. Insertion and deletion times were reduced by decreasing the number of accesses to the entire tree. Query processing accessed only the bookkeeping data structures and buckets. Non-leaf nodes of the tree were accessed during adjustments.
This test case measured the performance of a distributed static kd-tree implementation with one MPI process per KNL node and $>=64$ threads. These experiments are strong scaling, with the same dataset, and with increasing number of nodes and CPUs. The number of MPI ranks was varied from $16-256$. There are three values for number of threads - $64$, $128$ and $256$. The total number of cores ranges from $[1024-16384] $. Total number of threads ranges from $[1024-65536]$.
A uniform distribution with 1 billion 3D points sampled from $[1,1000000000] $ was used to test this configuration.
STL random distributions were used to generate uniform samples within fixed ranges.

\begin{figure}[!tbph]
\centering
\includegraphics[width=3.5in,height=2.5in]{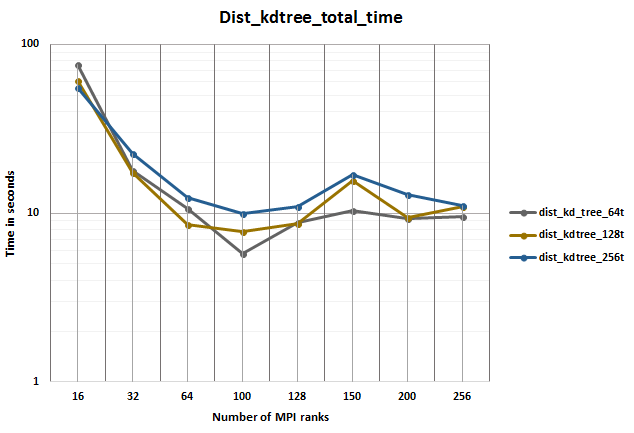}
\caption{\label{Fig:distkdtreetotal}Distributed KD-tree total time}
\end{figure}

The graph in figure~\ref{Fig:distkdtreetotal} shows the total time for all components, including load balancing and data transfer. The values on the y-axis of the graph are based on logarithmic scale. The graph shows some variation in scaling after $100$ MPI processes. The predominant cost in this region is data exchange compared to tree building. The time taken for data transfer depended on various factors, such as maximum number and size of messages, network latency, bandwidth and congestion.

\section{Applications~\label{appl}}
In this section we discuss a few applications that benefit from better partitions and data order in memory. The applications described here are from point location and general graph partitioning problems which demonstrate the versatility of this approach. 

\subsection{Point Location}
Search queries can be broadly classified into two categories based on their results : exact match and partial match. Exact match queries search for the exact data in the database. Partial match queries are a class of problems that include nearest neighbor searches and range queries. Out of these, this section deals with exact point location and K-nearest neighbor searches, where $K$ is a value selected by the application. The distance metric used here is Euclidean.

\subsubsection{Exact Point Location}

\begin{figure}[!tbph]
\centering
\includegraphics[width=8.5cm, height=7cm]{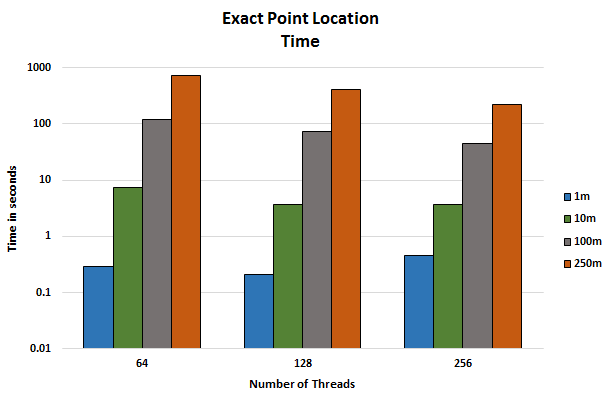}
\caption{\label{fig:pointloc} Point Location in Shared Memory}
\end{figure}

Input queries are presorted using their co-ordinates into bins~(equal to number of threads), where each bin covers the volume contained in a bounding box. Since the top $K1*K2*P$ nodes or bins are mapped to threads, point location queries can be executed in parallel. For each query a representative key is generated by bit interleaving the binary representations of the co-ordinates of a d-dimensional point. This key is searched for in a sorted list of buckets~(sorted using SFC keys of buckets) using binary search. Once a matching bucket is found, it is searched to locate the point. This method is a fast implementation that stores only buckets. But it works only with Morton SFC on uniform distributions in which the splitting hyperplanes cycle between the $d-1$ dimension planes in a fixed order and the splitting value is the midpoint along the $d^{th}$ dimension. For non-uniform distributions and Hilbert-like SFCs, non-terminal nodes have to be stored and point location will require tree traversals from sub tree roots to buckets. In both cases, the cost of point location is $O(logN)$ where $N$ is the number of buckets. The measured time in this section includes presorting and binning costs. Morton order was the SFC used in this experiment.
The tests were performed for points in 3D, with data sizes ranging from 1 million points to 250 million points. All tests were performed on a single KNL node with thread counts varying from 64-256. The graphs in figure~\ref{fig:pointloc} show the total time taken for exact point location. The y-axis has logarithmic values.

The algorithm we used for K-Nearest Neighbor~(K-NN)~\cite{Samet} search is similar to that used for exact point location.
Query points are presorted and binned based on the partitioning of $K1*K2*P$ nodes. Depending on the tree splitters and SFC, binary search on sorted buckets may be used to locate the SFC key generated by bit-interleaving point co-ordinates or top-down traversals may be used to locate buckets in sub trees. Once the point is located, the bucket containing the point, along with other buckets in the vicinity~($CUTOFF$) are searched for k-nearest neighbors. The number of buckets searched depends on the definitions of $BUCKETSIZE$ and $CUTOFF$. The largest sub tree containing the $CUTOFF$ volume of a point should be searched. In this experiment, we used Morton SFC and restricted $CUTOFF$ to one bucket before and after a bucket in the SFC.

\begin{figure}[!tbph]
\centering
\includegraphics[width=8.5cm, height=7cm]{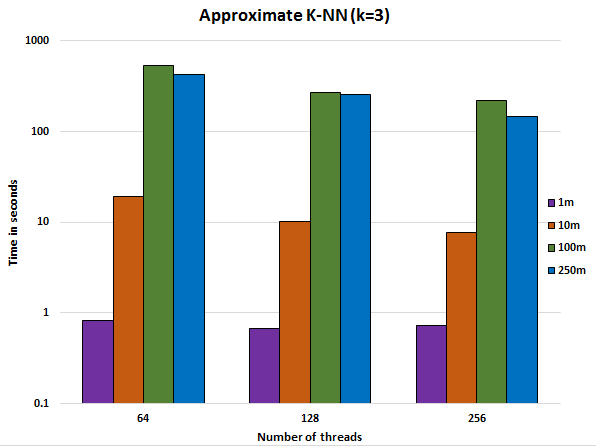}
	\caption{\label{fig:knn} Approximate K-NN in Shared Memory}
\end{figure}

The closest $K$ neighbors are chosen from all neighboring points within the $CUTOFF$ volume of a point. The test cases used here are for points in 3D. The input points are generated by sampling from within the kd-tree bounding box. The size of the input set is $100 million$. All experiments were conducted on a single KNL node with threads varying from $64-256$. The values for $CUTOFF$ and $K$ were $500000$ and $3$ respectively.

\subsection{General Graph Partitioning}
Many graph connectivity and traversal problems can be solved efficiently in parallel by using their linear algebra equivalents. There are libraries that offer good parallel solutions for dense and sparse linear computations such as ~\cite{blas},~\cite{linpack} which can be used to solve linear algebra formulations of graph problems. GraphBLAS~\cite{graphblas} is one such library that has implemented routines for common matrix computations in graph problems such as dense matrix-vector, dense matrix-matrix, sparse matrix-vector, sparse matrix-matrix and matrix inverse. In distributed graph problems, matrices and vectors should have good quality partitions for scalable performance. Besides Metis\cite{Metis} and Parmetis\cite{parmetis} the partitioner discussed in this paper can be used to partition dense and sparse matrices.
Metrics such as computational load per process/thread and inter process communication volume are used for comparing the quality of partitions. The entire computation is partitioned by partitioning the non-zero values in the sparse matrix and the dense vector. The row and column indices of the adjacency matrix are used as co-ordinates in 2 dimensional space. A non-zero at location $(i,j)$ is multiplied with a vector value at row index $j$. One can define an edge between a non-zero element in the adjacency matrix and a vector value. The non-zeros of the sparse matrix are partitioned using methods described earlier for static datasets. The dense vector is greedily partitioned into load balanced non-overlapping chunks~(\emph{owned}) across processes. Each process computes its required set of contiguous vector intervals from its sparse matrix partition. Vector intervals outside the range of a process's \emph{owned} chunk are referred to as \emph{dependent}. The \emph{dependent} vector intervals are replicated on processes. Every process performs the following operations in a distributed sparse-matrix dense vector multiplication :
\begin{itemize}
\item Compute local matrix-vector product
\item Reduce partial results from all processes with replicated vector intervals at owning processes. Scatter reduced vector sub intervals to replicated processes.
\end{itemize}
The inter process communication for reducing partial results and distributing subintervals is implemented using reduce-scatter routines provided by MPI. Each \emph{owned} chunk is the root of a communication tree with the replicated processes as leaves. Non-intersecting communication trees can perform reduce-scatter in parallel. The total inter process communication is reduced by minimizing replicated intervals. If the entire vector is replicated on all processes inter process communication volume is the vector size multiplied by the number of processes, which is the maximum communication volume for this problem. A combination of partitioning and replication is used to reduce communication volume. This problem can be reduced to one-dimensional range search on the dense vector which can be performed in constant time per query. For a fixed partitioning of the sparse matrix, the vector distribution that minimizes queries can be determined by computing a spanning set that covers the range of vector intervals. The communication graph can be visualised using a bipartite graph of vector interval queries~($X$) on the L.H.S and a set of processes~($R$) on the R.H.S. Edges are drawn between $X$ and the $R$s for a set of queries and vector distribution with weights on edges that are proportional to the cost of the communication. Memory accesses are assigned lower weights than data transfers over the network. The spanning set~($S\subseteq R$) is a non-overlapping distribution of vector intervals on processes that covers all queries by minimizing the total weight of communication edges. This spanning set~($S$) is included in the bi-partite communication graph by adding weighted edges from the L.H.S to the spanning set and from it to the R.H.S. Edges between $X$, $S$ and $R$ are the costs of reducing partial vector intervals on the spanning set and scattering results to processes. Iterative methods may be used to determine the spanning set that minimizes total edge weight from an initial set. Dense-matrix dense-vector multiplication algorithms have good solutions that minimize communication volume. For example, $P$ processes may be arranged in a two-dimensional mesh of $\sqrt P$ rows and columns, with the vector partitioned into $\sqrt P$ chunks along columns and replicated along $\sqrt P$ rows in each column. It is difficult to find such solutions for sparse-matrix dense-vector multiplications. Our implementation used the \emph{owned} chunks as the initial spanning set. We modified the spanning set once by assigning chunks to processes that have maximum overlap with their vector subintervals. In case of ties, the process with minimum id was chosen as \emph{owner}. Besides improving load balance and communication metrics, SFC orders improve spatial and temporal locality in cache accesses because geometry preserves the computation pattern of the problem.

These methods are applicable to large simulations which consist of several meshes or matrices. An example would be a multi-physics climate simulations consisting of adjacency matrix representations of atmosphere, ocean, ice and land meshes~\cite{climate}. Separate invocations of the SFC partitioner can be used to partition and load balance these 2 dimensional sparse matrices. If any of the meshes are adaptive they can be managed using general methods described for dynamic applications earlier in this paper.
Real world datasets~(graphs) were used as test cases for this section. This section has a set of empirical results for distributed sparse matrix dense vector multiplication. The metrics used for comparison are load balance~(average and maximum), number of messages~(MaxDegree) and communication volume~(MaxEdgeCut). SFC partitions are compared against a row-wise matrix decomposition wherein each process is assigned a fixed number of rows. The datasets were obtained from SNAP~\cite{snap}. They are Google, Orkut and Twitter social networks. The Google network was a square matrix of dimensions $916,428X916,428$ and $5,105,039$ non-zeros. The Orkut network had total size $3,072,441X3,072,441$ and $117,185,083$ non-zeros and the Twitter network~($41,652,230X41,652,230$) had $1,468,365,182$ non-zeros.

\begin{table*}[!tbph]
\centering
\begin{tabular}{|c|c|c|c|c|}
\hline
\#procs & AvgLoad & MaxLoad & MaxDegree & MaxEdgeCut \\
\hline
16 &	319064 & 322068	& 15 & 13034 \\
\hline
32 & 159532 &	162447	& 31 & 4387\\
\hline
64 &	79766	& 81936	& 63 & 1450\\
\hline
100 & 51050 & 53239 &	99 &	704\\
\hline
128 &	39883	& 41731 & 127	& 477\\
\hline
150 &	34033	&35561&	149 &	356\\
\hline
200& 25525 & 26842 &	199 & 225\\
\hline
256 &	19941 &	21059	& 255	&157\\
\hline
\end{tabular}
\vspace{1mm}
\caption{\label{tab:googlenet1}Empirical Measurements for GoogleNetwork Row-wise Partitions}
\end{table*}

\begin{table*}[!tbph]
\centering
\begin{tabular}{|c|c|c|c|c|c|}
\hline
\#procs & AvgLoad & MaxLoad & MaxDegree & MaxEdgeCut & Partitioning Time \\
\hline
16 & 319064 & 319065 &	3 & 1815 & 0.517941\\
\hline
32 &	159532	&159533 & 3 &982 & 0.487025\\
\hline
64 & 79766 & 79767 & 7 & 929 &	0.567527\\
\hline
100 &	51050 & 51051 &	10 & 1283 & 0.723744\\
\hline
128 &	39883&	39884&	10 &522&	0.742583\\
\hline
150&	34033&	34034&	16&	575&	1.10206\\
\hline
200&	25525&	25526&	19&	675&	1.05934\\
\hline
256&	19941&	19942&	22&	305&	1.03896\\
\hline
\end{tabular}
\vspace{1mm}
\caption{\label{tab:googlenet2}Empirical Measurements for GoogleNetwork SFC Partitions}
\end{table*}

\begin{table*}[!tbph]
\centering
\begin{tabular}{|c|c|c|c|c|}
\hline
\#procs & AvgLoad & MaxLoad & MaxDegree & MaxEdgeCut \\
\hline
32 & 3662033 &5976772& 31&84768\\
\hline
64& 1831016& 3666860& 63& 41534\\
\hline
100& 1171850&	2453572& 99& 25681\\
\hline
128 &915508 &2959793 &127 &19959\\
\hline
150& 781233 &2038979 &149 &16858\\
\hline
200 & 585925 &1414928 & 199 &12289\\
\hline
256 & 457754 & 1189407 & 255 &9356\\
\hline
\end{tabular}
\vspace{1mm}
\caption{\label{tab:orkutnet1}Empirical Measurements for OrkutNetwork Row-wise Partitions}
\end{table*}

\begin{table*}[!tbph]
\centering
\begin{tabular}{|c|c|c|c|c|c|}
\hline
\#procs & AvgLoad & MaxLoad & MaxDegree & MaxEdgeCut & Partitioning Time \\
\hline
32 &3662033 &3662034 &5	&14717 &8.45549\\
\hline
64 & 1831016 &1831017 &10 &19913 &4.50847\\
\hline
100 &1171850 &1171851 &11 &9922	& 3.46279\\
\hline
128 &915508 &915509 &11	& 15038 & 6.94193\\
\hline
150 & 781233 & 781234 & 23 & 5515 & 8.42405\\
\hline
200 &585925 &585926 & 23 & 13125 &7.5078\\
\hline
256 & 457754 & 457755 & 23 & 13345 &8.37234\\
\hline
\end{tabular}
\vspace{1mm}
\caption{\label{tab:orkutnet2}Empirical Measurements for OrkutNetwork SFC Partitions}
\end{table*}

\begin{table*}[!tbph]
\centering
\begin{tabular}{|c|c|c|c|c|}
\hline
\#procs & AvgLoad & MaxLoad & MaxDegree & MaxEdgeCut \\
\hline
32 & 45886411 &230950550 &31 &800810\\
\hline
64 &22943205 &150796780	& 63 &381688\\
\hline
100 &14683651 &119621190 &99 &240120\\
\hline
128 &11471602 &104492640 &127 &184585\\
\hline
150 &9789101 &95083723	&149 &158227\\
\hline
200 &7341825 &82417545&	199& 1146662\\
\hline
256 &5735801& 71120083&	255& 87908\\
\hline
\end{tabular}
\vspace{1mm}
\caption{\label{tab:twitternet1}Empirical Measurements for TwitterNetwork Row-wise Partitions}
\end{table*}

\begin{table*}[!tbph]
\centering
\begin{tabular}{|c|c|c|c|c|c|}
\hline
\#procs & AvgLoad & MaxLoad & MaxDegree & MaxEdgeCut & Partitioning Time \\
\hline
32 &45886411 &45886412& 5 &107437 &199.251\\
\hline
64 &22943205& 22943206& 12& 76513& 123.78\\
\hline
100 &14683651&	14683652& 15&	45321&	60.2393\\
\hline
128 &11471602&	11471603& 14&	46462&	58.7428\\
\hline
150 &9789101& 9789102&	17& 43742 &52.8559\\
\hline
200 &7341825& 7341826&	19& 33892& 55.4281\\
\hline
256 &5735801& 5735802& 27& 39742& 56.2482\\
\hline
\end{tabular}
\vspace{1mm}
\caption{\label{tab:twitternet2}Empirical Measurements for TwitterNetwork SFC Partitions}
\end{table*}

The observations in tables~\ref{tab:googlenet1},~\ref{tab:googlenet2},~\ref{tab:orkutnet1},~\ref{tab:orkutnet2},~\ref{tab:twitternet1} and ~\ref{tab:twitternet2} show the benefits of using SFC partitions compared to row-wise decompositions. SFC partitions have consistently lower degrees and edge-cuts, which implies fewer inter process messages and reduced communication volume during reduce-scatter.  

\section{Conclusions and Future Work}
In this paper, we discussed the performance of a parallel geometric partitioner that scales well on many-core processors. We also developed general methods for partitioning and load balancing dynamic applications such as parallel query processing and Delaunay refined meshes. This partitioner produces better quality partitions compared to other geometric partitioners. These partitions are comparable to those produced by linear optimization methods. Our efforts extended the scope of geometric partitioners beyond structured meshes. We also defined amortized load balancing techniques for dynamic data. We demonstrated the wide scope of this partitioner using applications from different domains of computer science. A fast highly parallel geometric partitioner would benefit the HPC community. The observations in this paper are from a cluster built from a many-core processor that can also function as a co-processor. Therefore the algorithms can be ported to GPUs, if needed. Other methods for parallel partitioning have higher inter process communication, are not scalable on new many-core architectures and are not suitable for incremental load balancing.
As part of future work, we would like to run some real-world graph processing algorithms using these partitions. We would also like to use this method to partition large geospatial data sets such as those derived from environmental monitoring.

\bibliographystyle{IEEEtran}
\bibliography{thesisrefs} 

\end{document}